\documentclass[onecolumn]{aastex631}
\usepackage{natbib}
\usepackage{amsmath}
\usepackage{graphicx}
\usepackage{epstopdf}
\usepackage{amssymb}
\usepackage{ragged2e}
\usepackage{float}
\usepackage{soul}
\usepackage{subcaption}
\usepackage{graphicx}

\shorttitle{OGLE-IV Period-Luminosity relation}
\shortauthors{Mu\~noz et al.}

\begin{document}
%\linenumbers 

\title{OGLE-IV Period-Luminosity relation of the LMC: an analysis using mean and median magnitudes}

\correspondingauthor{Alejandro Garc\'ia-Varela}
\email{josegarc@uniandes.edu.co}

\author[0000-0002-0355-0900]{Jaime  Mu\~noz}
\affiliation{Universidad de los Andes, Facultad de Ingenier{\'i}a \\   Cra. 1 Este No. 19A - 40, Bloque ML \\ Bogot\'a, Colombia}

\author[0000-0001-8351-0628]{Alejandro Garc\'ia-Varela}
\author[0009-0007-1937-7675]{Santiago Henao-Castellanos}
\author[0000-0003-1480-8556]{Beatriz Sabogal}
\affiliation{Universidad de los Andes, Departamento de F{\'i}sica \\  Cra. 1 No. 18A-10, Bloque Ip, A.A. 4976 \\ Bogot\'a, Colombia}

\author[0000-0002-2492-4422]{Luis Felipe Giraldo}
\affiliation{Universidad de los Andes, Departamento de Ingenier{\'i}a Biom{\'e}dica \\  Cra. 1 Este No. 19A - 40, Bloque ML \\ Bogot\'a, Colombia}

\author{Jorge Mart\'inez\textsuperscript{$\dagger$}}
\affiliation{Universidad de los Andes, Departamento de Ingenier{\'i}a Industrial \\  Cra. 1 Este No. 19A - 40, Bloque ML \\ Bogot\'a, Colombia}
\altaffiliation{\textsuperscript{$\dagger$}\textnormal{This study is inspired by the ideas and suggestions of professor Jorge Mart\'inez Collantes, who sadly passed away in June 2017. 
We honor his memory.}}

\begin{abstract}

The Period-Luminosity (PL) relation for Cepheid variable stars in the Large Magellanic Cloud (LMC) is crucial for distance measurements in astronomy. This study analyzes the impact of using the median rather than the mean on the PL relation's slope and zero point. It also examines the persistence of the break at approximately 10 days and addresses specification issues in the PL relation model. Using $VI$-band median and mean magnitudes from the OGLE-IV survey, corrected for extinction, we fit the PL relation employing robust $MM$-regression, which features a high breakdown point and robust standard errors. Statistical tests and residual analysis are conducted to identify and correct model deficiencies. Our findings indicate a significant change in the PL relation for Cepheids with periods of 10 days or longer, regardless of whether median or mean magnitudes are used. A bias in the zero point and slope estimators is observed when using median magnitudes instead of mean magnitudes, especially in the $V$-band. By identifying and correcting regression issues and considering the period break, our estimators for slope and zero point are more accurate for distance calculations.
Comparative analysis of the models for each band quantifies the bias introduced by using median magnitudes, highlighting the importance of considering the Cepheids' period for accurate location measure results, similar to those obtained using mean magnitudes.
\end{abstract}


\section{Introduction}

The Period-Luminosity (PL) relation, first identified by \citet{HL}, revealed that the maximum and minimum brightness of 25 Cepheids in the SMC can be represented by two approximately parallel lines when plotted against the logarithm of their respective periods. 
Recognizing the potential of the PL relation for distance determination, subsequent studies focused on exploring various methodologies to identify a single statistical location measurement instead of relying on the two extreme values simultaneously. 
For example, \citet{Kr} converted magnitudes to an intensity scale, computed the median of intensities, and then converted this value to the magnitude scale. 
On the contrary, \citet{ST} utilized the maximum brightness of Cepheids to estimate the distance moduli of galaxies. 
Subsequent research on RR Lyr{\ae} and Cepheid stars highlighted the importance of accurately determining mean magnitudes since these radial pulsators are more commonly observed during their bright phase than during their faint one. 
To address this observational bias, \citet{Saha} proposed the concept of phase-weighted intensity-average magnitude, which considers the period of the star. 

For over two decades, the scientific community has widely employed the intensity-averaged magnitude as a reliable and robust measure of location. 
This is obtained by fitting the light curve with a truncated Fourier series \citep{Ngeow2003},
computing its mean intensity, and then converting it to a magnitude scale.
This method is applicable regardless of the distribution of a sufficient number of points on the light curve. 
\citet{Tur} showed that for uniformly sampled Cepheids, the intensity-averaged magnitude and the phase-weighted mean magnitude coincide. 
A summary of several research studies conducted in recent decades that have used intensity-averaged magnitude (mean magnitude, hereafter), is presented following. 
\citet{U99} derived $VI-$band PL relations, based on exceptionally well-defined hundreds of Cepheid light curves from the LMC observed for several hundred epochs during the second phase of the OGLE project. 
These relations constituted the fundamental calibrators for the distance measurements of galaxies reported by \citet{Freed}. 
In this study, mean magnitudes were utilized to calibrate the PL relations of Cepheids observed with the \textit{Hubble Space Telescope} in 31 distant galaxies. 
\citet{NGC300} calculated the mean magnitudes of the Cepheids in NGC 300 by fitting a second- and fifth-order Fourier series to the light curves. 
\citet{Sos2010} reported the OGLE-III SMC catalog of variable stars, deriving PL relations for thousands of Cepheids using mean magnitudes. 
Recently, \citet{Rip2023} presented a Gaia DR3 catalog containing thousands of Cepheids of all types, accompanied by PL relations derived from mean magnitudes. While the intensity-average method and its phase-weighted counterpart are effective when applied to datasets with sufficient sample sizes, challenges arise when dealing with data coming from sparse and noisy photometry. 
To address this issue, templates have been developed in order to reconstruct the shape of the light curve using only the period information. 
This was first done by \cite{Stetson1996}, who derived first-order Fourier coefficients for Cepheids in IC 4182. 
Later, higher-order Fourier templates for MW, LMC, and SMC Cepheids were derived in the NIR by \cite{Soszynski2005}, 
and in the $VI$-bands by \cite{Tanvir2005}, who also reduced the coefficients by principal component analysis.
More recently, \cite{Inno2015} proposed periodic Gaussian functions as a replacement for the Fourier basis.

    Additionally, the single-banded magnitude is not the only choice for the dependent variable in the PL relation. 
    Wesenheit indices, defined as a combination of magnitudes and colors, have been used in recent decades through Period-Wesenheit (PW) relations.
    These PW relations show lower dispersion than their magnitude counterparts \citep{Caputo2000, Bono2010}, 
    because the inclusion of color and total-to-selective extinction partially mitigates the effect of reddening and the temperature width of the instability strip \citep{Madore1991}; however, an extinction map is still needed for galaxies with complex dust structures like the LMC. Furthermore, the PW relations are less influenced by metallicity compared to the PL relation \citep{deSomma2022}. \cite{Anderson2016} demonstrated that the pulsation properties of classical Cepheids are affected by their evolutionary rotation, which in turn impacts the PL relations. However, the PW relations may be less susceptible to these effects due to their incorporation of color information.

For an updated review of Cepheids along with a discussion of the advantages and disadvantages of using the PL, PW, and Period-Color-Luminosity relations to determine distances, we recommend consulting the work by \cite{Bono2024}. Additionally, for a recent account on the utilization of PW relations in the extragalactic distance scale, we direct your attention to the studies by \citeauthor{Riess2019} \citeyearpar{Riess2019,Riess2023}.

One of the aspects of the PL relation that  
requires further understanding is the presence of a break in the slope occurring around 10 days, a phenomenon first identified by Kukarkin \citep{Fer69}, subsequently confirmed by \citet{Tamm2003} and later corroborated by \citet{Kanbur2004}, \citet{Kanbur2007}, \citet{Koen2007} and \citet{Ngeow2012}.

A theoretical explanation for this phenomenon was proposed by \citet{Kanbur2006}, who examined whether short-period Cepheids of the LMC ($P<10$ days) show an almost continuous interaction between the hydrogen ionization front and the photosphere throughout their pulsation cycle; in contrast to long-period Cepheids ($P>10$ days) where the photosphere only interacts at the time of maximum light. Another possible explanation for this phenomenon could be the influence over the PL relation of large amplitude Cepheids showing the Hertzsprung progression near a period of $10$ days \citep{GV2016}. Despite these possible explanations, more research is needed to understand this break and its dependence on metallicity. However, due to the limited number of known long-period Cepheids in the Milky Way, LMC, and SMC galaxies, it has not been possible to achieve a high level of confidence in the calibration of the PL relation in the bright regime.

Considering the absence of recent studies employing location measures other than those using mean magnitude obtained by the Fourier series or the template fitting methods, we aim to investigate the median of the magnitude distribution as an alternative location measure because of its robustness and high breakdown point.
We want to explore its impact on the slope and zero point of the LMC PL relations of OGLE-IV Cepheids when employing median magnitudes instead of mean magnitudes. 
In addition, when adopting the median as the location measure, our objective is to investigate whether or not the observed break in the optical PL relation around 10 days is mitigated. 

The required analysis to address these inquiries involves several steps. First, we preprocess the data by applying the extinction correction method, as explained in Section 2. Next, we perform the robust regression analysis for the PL relation and compare the models on each band using the median and mean magnitudes described in Section 3. In Section 4, the results obtained are presented and their implications are discussed. Finally, Section 5 provides a summary of the main conclusions.

\section{Data preprocessing}

The data used in this analysis comes from the OGLE-IV project \citep{ogleiv2015}, 
which has a considerably wider sky coverage at the outskirts of the LMC than its predecessor.
The OGLE Collection of Variables Stars (OCVS) provides $VI-$band light curves publicly available from their \mbox{website}\footnote{
    \url{www.astrouw.edu.pl/ogle/ogle4/OCVS/}
}, and they report 4709 LMC Classical Cepheids, 
with 2477 of them classified as fundamental overtone pulsators \citep{OGLEClassicalCepheids},
the main target of the PL relation.

In order to rule out extinction effects in our analysis, a reddening map must be used to correct the Cepheid magnitudes.
The full-sky dust extinction maps of \citet{Schlegel1998} (SFD) cannot be used inside the LMC, because their dust temperature structure is not resolved,
resulting in an overestimation of the extinction.
More recent observations have resolved the dust distribution on the LMC \citep{Utomo2019}, 
but these results are difficult to convert into reddening values due to poor knowledge of the dust size distribution and its emissivity properties \citep{Bell2020}.
In this study, maps made with indirect extinction measurements were preferred. 
The most recent reddening maps of the LMC are the ones by \citet{Bell2022} and \citet{Chen2022},
using spectral energy distribution fitting. 
These maps use data from the SMASH and VMC surveys, which do not have as much sky coverage as OGLE-IV.

Reddening maps with similar sky coverage as our star sample are, however, available. 
\citet{Skowron2021} produced an extinction map by correcting the reddening of the Red Clump (RC) on most of the Magellanic Clouds OGLE-IV fields.
The resolution bins of the map depend on the amount of RC stars present in each field, 
to preserve the statistical accuracy on the extinction values,
which causes the resolution of the map to go from $\sim 3.3''$ in the LMC center to $\sim73''$ in the outskirts. 
This is not a problem, as the distribution of Cepheids have a similar trend, 
with only three Classical Cepheids in our sample being outside the mapped area.
For this reason, we selected this map for the extinction correction.

The color excess is reported as $E(V-I)_{-\sigma_1}^{+\sigma_2}$, with the $50\pm34.1\%$ percentiles used to calculate the lower $\sigma_1$ and upper $\sigma_2$ uncertainties,
equivalent to a $\pm1\sigma$ on a normal distribution \citep{Skowron2021}.
The \texttt{.fits} image with this data was downloaded from the OGLE webpage\footnote{\url{ogle.astrouw.edu.pl/ms_ext/skowron2020_all.fits}}.
We converted the sky coordinates of each star into the pixel values of the image and used linear interpolation to obtain the color excess of each star.
We used $R_{VI} = 1.67$ \citep{Skowron2021} to turn these color excess values to $A_V$ and $A_I$ extinction coefficients.

\begin{figure}[h]
    \centering 
    \vspace{5mm}
    \includegraphics[width=0.45\textwidth]{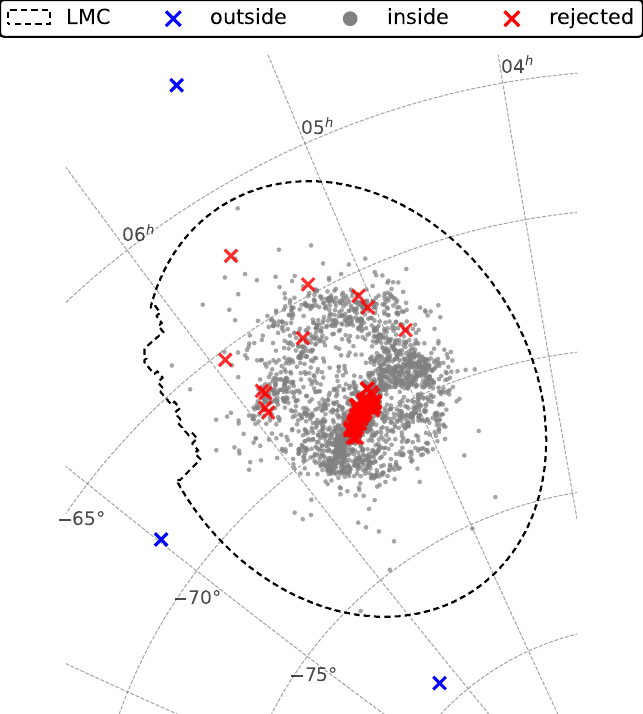}
    \caption{Sky coordinates on the OGLE-IV fundamental mode classical Cepheids in the LMC, as grey dots. The \citet{Skowron2021} extinction map limits are shown in a dashed line, and the rejected stars are marked with crosses; the ones rejected because of the uncertainty of their color excesses in red, and those that lie outside of the extinction maps in blue.}
     \label{fig:LMC_selection}
\end{figure}

In some crowded regions, however, a secondary RC can be found, which widens the distribution used to calculate the reddening,
and contaminates it with younger stars.
This causes the lower bound $E(V-I)-\sigma_1$ to fall below zero in certain areas.
As a negative color excess is unphysical, we removed 96 stars that presented this problem.
A finding chart of this deletion is presented in Figure \ref{fig:LMC_selection}.

After the extinction correction, the intensity-averaged magnitude and the median magnitude were computed for each star using the full light curve.
The OCVS lacked light curve data for 180 of the listed Cepheids on the $V-$band and 46 on the $I-$band,
leaving our final sample selection without missing values on 2193 and 2326 stars, respectively.  

\section{Methodology}

The PL relations were fitted using Ordinary Least Squares (OLS) with the mean and median magnitudes in the $VI-$bands. This method assumptions are: constant variance in the errors (homoscedasticity), no influential points in the data, and assessing if the errors follow a normal distribution, although in practice the errors must not have strong departures from the normal distribution. %\citep{Mont}.
Upon examining the OGLE-IV LMC Cepheid data and analyzing the models' residuals using measures of influence  \citep{Mont}, we identified influential points that violate the OLS assumptions. To deal with influential data points, we use a robust method to fit the PL relations. In the context of this analysis, the chosen robust method can also handle heteroscedasticity and uncorrelated errors. Due to that, the PL relations were fitted using $MM$-regressions, that have a high breakdown point (BDP) of $50\%$ that improved the robustness of this regression, with $95\%$ asymptotic efficiency for normal errors. The BDP represents the maximum proportion of anomalous data required to make the estimators useless \citep{RobStat}.  

Therefore, a high BDP such as $50\%$ of the $MM$-regression means that even if half of the data are contaminated; for example, because there are influential points, the model estimators are still useful. We chose $MM$-regression over other robust methods because it effectively combines robustness and high asymptotic efficiency, allowing it to handle influential points without losing precision, even extreme ones. Additionally, when robust standard errors are used, heteroscedasticity does not affect this regression and its estimators are not biased \citep{Feigelson2021}. Besides, research by \cite{yu2017robust} demonstrates $MM$-regresion     outperforms other methods in terms of performance, and statistical software such as \texttt{R}   has packages that directly implement this regression \citep{RS}. %This improved $MM$-regression was implemented%
One of these packages is  \textbf{\texttt{Robustbase}}  \citep{Rbase}, it implements $MM$-regression, with Tukey's bi-square $\rho$ function \citep{Drap}. In addition, \texttt{Robustbase} uses the robust coefficient of determination given by \citet{Renaud}, and also uses robust standard errors \citep{Croux} that require neither homoscedasticity, nor uncorrelated errors, nor symmetry in the error distribution, as was explained by \cite{RobStat}.
  
After fitting a PL relation model, it is important to analyze the model residuals to look for evidence of problems related to the regression functional form; for instance, the presence of more or fewer variables than the model requires (specification problems).  
To do that we perform the \textit{robust slope test}, followed by the \textit{Ramsey-Reset} specification test \citep{God}.
Both tests are explained in Appendix A. 
Additionally, the model's residual plots were analyzed to look for any pattern (i.e. linear or quadratic) that disrupts the random behavior
expected in the residuals. Because these tests and the residual plots showed problems, the models were adjusted using dummy variables.

To compare the resulting models,  hypothesis testing using dummy variables, as explained by \cite{Drap} and \cite{Mont} was performed to quantify the impact of using the median magnitude in the PL relation on each band. More details about this test can be found in Appendix A.

\section{Results and discussion}

In this section, we obtained the $I-$band PL relations using mean ($\mathbf{\boldsymbol{I_0}}$) or median ($\mathbf{\boldsymbol{I_0^M}}$) magnitudes corrected for extinction, in linear regression models. 
We will use a bold font to denote vector quantities. The PL relation can be modelled by $\mathbf{\boldsymbol{I}}=\beta_1+\beta_2 \ \boldsymbol{{\log 
 P}}+\boldsymbol{\varepsilon}$, where 
 $\boldsymbol{\log P}$ is the logarithm of the period pulsation of a Cepheid (in days) and $\boldsymbol{\varepsilon}$ is the error in the model. 
The model was first adjusted
by OLS as \citet{U2000} did, but after applying influence measures to the model residuals \citep{Mont}, influential  
points were found. 
Since there were no non-statistical reasons to discard these influential points, a robust regression was needed to deal with them because their presence violated the assumptions of OLS. 
As a result, the $MM$-regression was used.

\begin{figure*}
    \begin{subfigure}{0.49\textwidth}
        \centering
        \includegraphics[width=\linewidth]{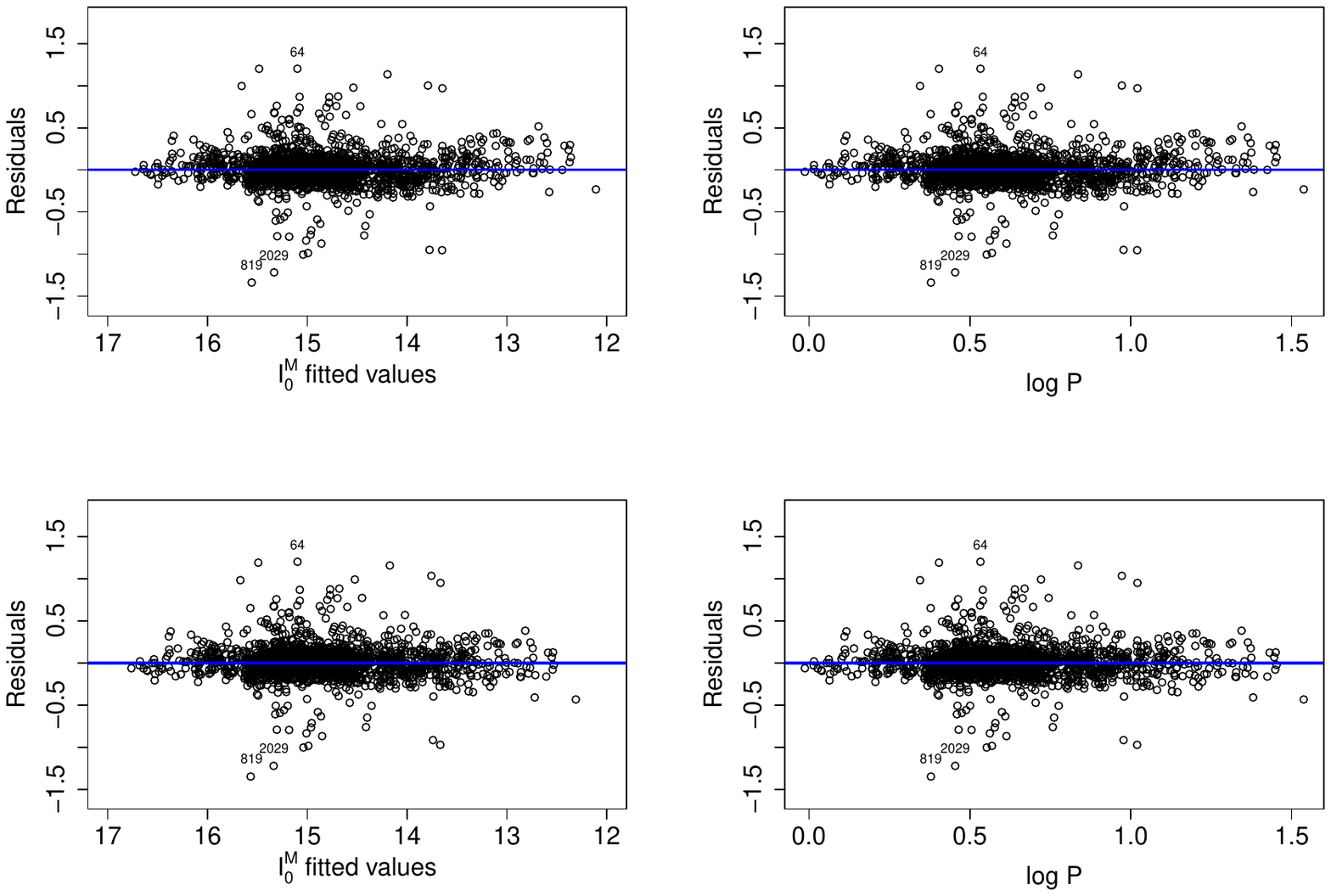}
        \caption{Residual plots}
        \label{fig:2a}
    \end{subfigure}%
    \hfill
    \begin{subfigure}{0.49\textwidth}
        \centering
        \includegraphics[width=\linewidth]{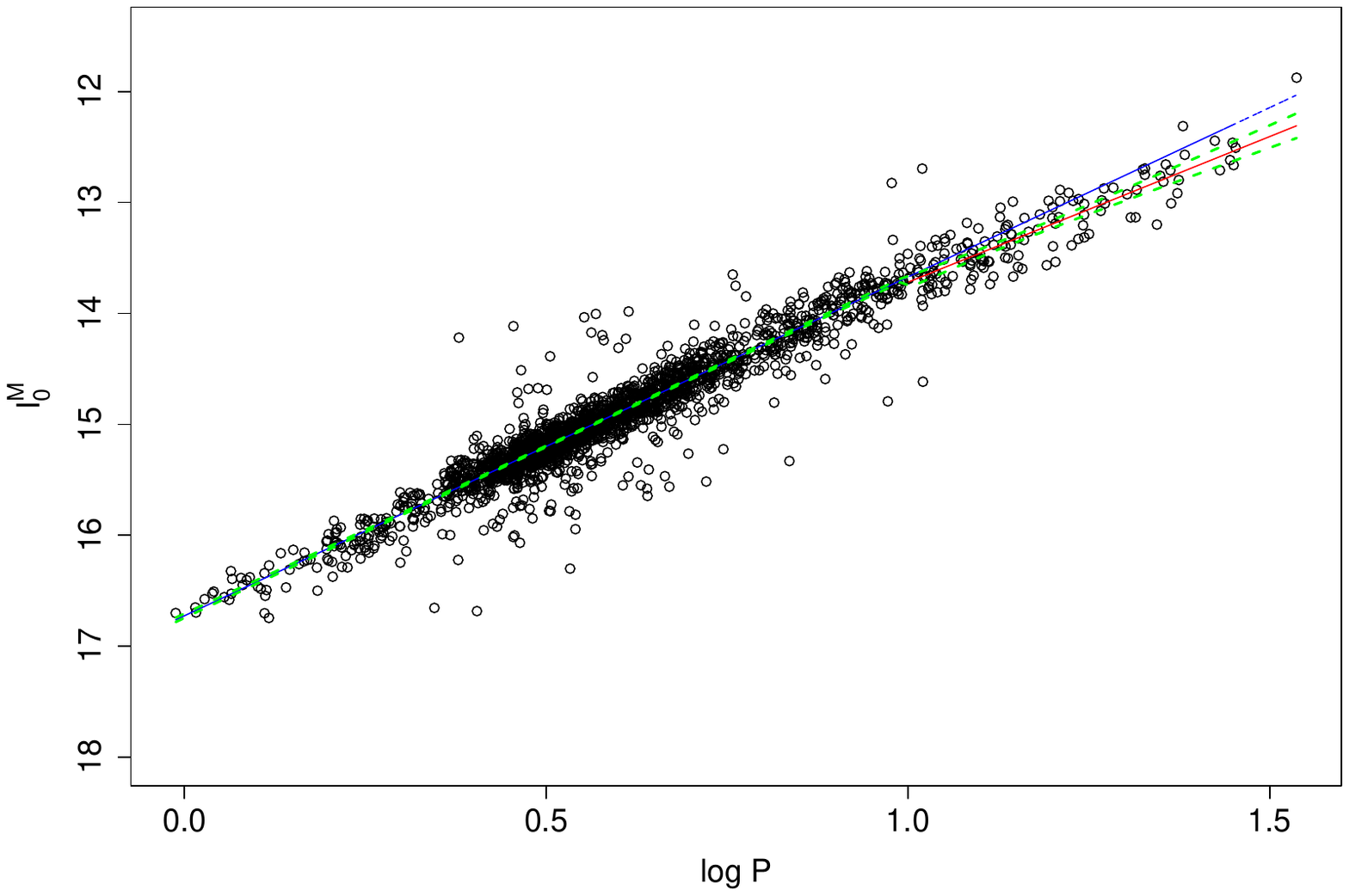}
        \caption{PL relation fitted}
        \label{fig:2b}
    \end{subfigure}
    \caption{PL relation fitted in $\mathbf{\boldsymbol{I_0}^M}$ by $MM$-regression. Plot (a): Checking the model residuals implies statistically plotting the residuals vs the fitted values (upper left panel) and the residuals vs the logarithm of the period (upper right panel). The blue line helps to notice the residual tendency changes. 
   Observations 64, 819, and 2029 have the highest residuals and correspond to Cepheids OGLE-LMC-CEP-0130, OGLE-LMC-CEP-1583, and OGLE-LMC-CEP-4023 respectively. Plot (b): The regression given by the Equation \eqref{eq10} is shown in a blue line for Cepheids with a $\log \ P<1$ and in a red line for Cepheids with a $\log \ P \geq{1}$. The blue line has been projected further with $\log \ P \geq{1}$ to enhance the visibility of the change when compared to the red line.  The area between the green dashed lines shows the $95\%$ confidence interval for $\mathbf{\boldsymbol{I_0^M}}$ expected value. 
    This area increases when $\log \ P \geq{1}$ because Cepheids having these periods are more scattered around the regression line.}
    \label{fig:2}
\end{figure*}

The first model we fitted was $\boldsymbol{I_0^M}=\beta_1+\beta_2\ \boldsymbol{\log P}+\boldsymbol{\varepsilon}$. 
To check the model, some tests were performed. 
First of all the \textit{robust slope test}, obtaining a $p$-value less than $0.0001$, so it is significant. 
To continue checking the model, the specification test was also performed. 
The \textit{Ramesey-Reset test} was found significant because its $p$-value was less than $0.0001$, which means there is strong evidence of specification problems in the model. 
Furthermore, examining the residuals for the model in Figure \ref{fig:2}(\subref{fig:2a}), the upper left panel depicts a change in the residual model trend for magnitudes ranging from 13.3 to about 12.  In addition, the upper right panel plot shows a break in the residuals when $\log \ P \geq{1}$, which is evident for residuals above the blue trend line. This behavior indicates a trend shift when Cepheids have periods longer or equal to 10 days.
This issue indicates specification problems in the model. 
The heteroscedasticity shown in Figure \ref{fig:2}(\subref{fig:2a}) is not a problem in $MM$-regression because of the robust standard errors used in the model \citep{Croux}.

Because of the specification problem detected in the model, a new one was fitted $\mathbf{\boldsymbol{I_0^M}}=\beta_1+\beta_2\ \boldsymbol{\delta}+\beta_3\ \boldsymbol{{\log P}+\beta_4\ \boldsymbol{\delta}\ \boldsymbol{\log P}}+\boldsymbol{\varepsilon}$. 
This new model introduced a dummy variable ($\delta$) to adjust the zero point and the slope when $\log \ P \geq{1}$. 
As a result, $\delta=1$ if $\log \ P \geq{1}$ otherwise $\delta=0$, as explained in Appendix A. 
Once again, the \textit{robust Slope test} was performed on the new model, and its obtained $p$-value was less than $0.0001$, indicating that the new model is globally significant. 
Therefore, the \textit{Ramsey-Reset test} was also performed to check for specification issues in the new model. This test is not statistically significant, as indicated by the $p$-value of $0.2654$.
The residual plots of the new model, shown in Figure \ref{fig:2}(\subref{fig:2a}) (bottom panel), are consistent with the specification test result since the trend shift found before disappeared with the adjustments incorporated in the new model. There is no visible change in the residual trend. 

The values of $\beta_1$ (zero-point), $\beta_2$ (adjustment to zero point when $\log \ P \geq{1}$), $\beta_3$ (slope),  and $\beta_4$ (slope adjustment when $\log \ P \geq{1}$) with their respective confidence intervals at $95\%$ obtained are reported in Equation (\ref{eq10}), and the regression plot is shown in the Figure \ref{fig:2}(\subref{fig:2b}).
\begin{equation}
  \begin{split}
    \mathbf{\boldsymbol{I_0^M}} & =(16.7262\pm0.0192)-(0.3815\pm0.3133) \ \boldsymbol{\delta}  \\ 
    & -(3.0540\pm0.0342)\ \boldsymbol{\log P}  \\ 
    & +(0.4283\pm0.2733) \ \boldsymbol{\delta \log \ P}+ {\boldsymbol{\hat{\varepsilon}.}} 
    \label{eq10}
  \end{split}
\end{equation}

A similar analysis for the dependent variable $\mathbf{\boldsymbol{I_0}}$ was made. 
 The OLS method was used to fit the PL relation, and influential points were detected. Consequently, the robust $MM$-regression described previously was utilized again.
Starting with the  \textit{robust slope test}, the obtained $p$-value was less than $0.0001$ indicating the model is globally significant. 
Continuing with the \textit{Ramsey-Reset test}, its $p$-value was $0.0014$ indicating that there is enough evidence of specification problems in the model at $5\%$ significance level. 
The break in the model residuals (trend shift) can be seen in Appendix B, Figure \ref{fig:3}(\subref{fig:3a}) (upper right panel). 
This break is also related to Cepheids with periods of 10 days or longer. 
Hence,  once again the adjustments to the zero point and the slope were needed to improve the model. 
 The latter led to the new model $\mathbf{\boldsymbol{I_0}}=\beta_1+\beta_2\ \boldsymbol{\delta}+\beta_3\ \boldsymbol{{\log P}+\beta_4\ \boldsymbol{\delta}\ \boldsymbol{\log P}}+\boldsymbol{\varepsilon}$. 
The new model \textit{robust slope test} is significant since its $p$-value was less than $0.0001$ and there is not enough evidence of specification problems since \textit{Ramsey-Reset test} $p$-value was equal to $0.3038$. 
The new model residual plots display heteroscedasticity, which is not an issue as explained before. There is also no indication of a break or change in the model residual trend (Figure \ref{fig:3}(\subref{fig:3a}), bottom panel, Appendix B). 
 
To summarize, no evidence was found by the specification test or the residual plots which may suggest issues in the model. 
Because all the model parameters were significant at $5\%$ significance level, the regression plot is shown in Appendix B, Figure \ref{fig:3}(\subref{fig:3b}) and the following values of $\beta$s and their respective confidence intervals at $95\%$ are reported as follows in Equation (\ref{eq11}):

\begin{equation}
\begin{split}
\mathbf{\boldsymbol{I_0}} &=(16.6740\pm0.0198)-(0.4127\pm0.3071) \boldsymbol{\delta}  \\
& -(2.9767\pm0.0350) \ \boldsymbol{\log P} \\ 
& +(0.4237\pm0.2706) \ \boldsymbol{\delta \log P}+ \boldsymbol{\hat{\varepsilon}.}   
\label{eq11}
\end{split}
\end{equation} 

The models given by Equations (\ref{eq10}) and (\ref{eq11}) have robust residual standard errors of 0.1173 and 0.1218, explaining $96.59 \%$ and $96.22 \%$ of the variability of $\mathbf{\boldsymbol{I_0^M}}$ or $\mathbf{\boldsymbol{I_0}}$, respectively, employing the adjusted robust coefficient of determination. 
The standard errors of the parameter estimators for $\boldsymbol{\delta}$ and $\boldsymbol{\delta \log P}$ were larger indicating more variability in Cepheids with $\log P > 1$ because these stars are more scattered around the regression line. 
Moreover, there are only 141 out of 2326 Cepheids having these periods. 

The same approach was used, to analyze the PL relation in $V-$band, which can be modeled by $\mathbf{\boldsymbol{V}}=\beta_1+\beta_2 \ \boldsymbol{{\log 
 P}}+\boldsymbol{\varepsilon}$, 
where $\boldsymbol{V}$ can be the median or mean, $\mathbf{\boldsymbol{V_0^M}}$ or $\mathbf{\boldsymbol{V_0}}$ respectively, $\boldsymbol{\log P}$ is the logarithm of the period pulsation of a Cepheid (in days) and $\boldsymbol{\varepsilon}$ is the error in the model. 
The PL relation was fitted for $\boldsymbol{{V_0}^M}$ by OLS. Because
 influential points were detected, the PL relation was refitted using the robust $MM$-regression already described. 
 To check the model, the first test performed was \textit{robust slope test}, whose $p$-value was less than $0.0001$; therefore, the model is globally significant. 
 Then, the  \textit{Ramsey-Reset test}  was performed getting also a $p$-value that was less than $0.0001$, thus, there is enough evidence of specification problems. 
 The model residual plots (Figure \ref{fig:4}(\subref{fig:4a}), upper panel, Appendix B) are concordant with the specification test because they show the same break already explained before. 
  The specification problems detected led to a new model that considered the behavior of Cepheids with larger periods. 
 The same approach applied to $I-$band was used to specify the new model $\mathbf{\boldsymbol{V_0}^M}=\beta_1+\beta_2\ \boldsymbol{\delta}+\beta_3\ \boldsymbol{{\log P}+\beta_4\ \boldsymbol{\delta}\ \boldsymbol{\log P}}+\boldsymbol{\varepsilon}$. 
 Because the  \textit{robust slope test}'s $p$-value was less than $0.0001$, the new model is globally significant. 
 Moreover, all the resulting model parameters are significant at $5\%$ significance level. 
However,  \textit{Ramsey-Reset test} was conducted and still being significant at $5\%$ significance level with a    
      $p$-value equal to 0.0237 although there is no evidence of problems in the model residual plots (Figure \ref{fig:4}(\subref{fig:4a}), bottom panel, Appendix B). 
Hence, the resulting model Equation (\ref{eq12}) is reported with $95\%$ confidence intervals for its $\beta$s and the regression plot in Figure \ref{fig:4}(\subref{fig:4b}), Appendix B. 

\begin{equation}
\begin{split}
\mathbf{\boldsymbol{V_0^M}} &=(17.3275\pm0.0263)-(0.6711\pm0.4461) \ \boldsymbol{\delta}  \\ 
 &-(2.9055\pm0.0474)\ \boldsymbol{\log P} \\ 
 &+(0.7153\pm0.3839) \ \boldsymbol{\delta \log \ P}+ \boldsymbol{\hat{\varepsilon}.} 
\label{eq12}
\end{split}
\end{equation}

Similarly, the analysis for $\mathbf{\boldsymbol{V_0}}$ revealed the presence of influential points when fitting the PL relation using OLS. Therefore, the previously described robust regression method was utilized once again.
 As done before, the first test conducted was the  \textit{robust slope test}, its $p$-value was less than $0.0001$ indicating that the model is globally significant. 
Then, specification problems were detected with the  \textit{Ramsey-Rest test} since its $p$-value was less than $0.0006$. 
 The model residual plots (Figure \ref{fig:5}(\subref{fig:5a}), upper panel, Appendix B) show the same issue, a change in the tendency of the residuals, especially noticeable in the residuals vs $\log P$ plot at the upper right panel. 
One more time, a new model was implemented to address this behavior and resolve the specification problem. As done before, the new model was: $\mathbf{\boldsymbol{V_0}}=\beta_1+\beta_2\ \boldsymbol{\delta}+\beta_3\ \boldsymbol{{\log P}+\beta_4\ \boldsymbol{\delta}\ \boldsymbol{\log P}}+\boldsymbol{\varepsilon}$. 
 To check this new model the \textit{Robust slope test} and the \textit{Ramsey-Reset test} were performed. 
 These tests $p$-values were less than 0.0001 and equal to 0.6021 respectively. 
 These results indicate that the model is globally significant, and there is insufficient evidence to suggest it has specification problems.
The new model residual plots (Figure \ref{fig:5}(\subref{fig:5a}), bottom panel, Appendix B) do not show evidence that indicates that the model is inadequate. 
Therefore, because all the model parameters were significant, the resulting model's $\beta$s with their respective $95\%$ confidence intervals are reported as follows in Equation (\ref{eq13}) and the regression plot in Figure \ref{fig:5}(\subref{fig:5b}), Appendix B.

\begin{equation}
  \begin{split}
\mathbf{\boldsymbol{V_0}} & =(17.2011\pm0.0268)-(0.4153\pm0.4020) \ \boldsymbol{\delta} \\ 
& -(2.7623\pm0.0478)\ \boldsymbol{\log P} \\ 
& +(0.4393\pm0.3479) \ \boldsymbol{\delta \log \ P}+ \boldsymbol{\hat{\varepsilon}.} 
\label{eq13}
  \end{split}
\end{equation}

The models given by Equations (\ref{eq12}) and (\ref{eq13}) have robust residual standard errors of 0.1591 and 0.1661, explaining the $93.11\%$ and $92.26\%$ of the variability of $\boldsymbol{{V_0}^M}$ and $\boldsymbol{V_0}$ respectively, using the adjusted robust coefficient of determination. 
Moreover, the standard errors of the parameter estimators for $\boldsymbol{\delta}$ and $\boldsymbol{\delta \log P}$ were larger indicating more variability in Cepheids with $\log P > 1$ because these stars are more scattered around the regression line. 
Besides, there are only 139 out of 2193 having periods. \\

\begin{table*}[t]
\centering
\caption{Bias in the PL relation models when using the median and the mean on each band.}
\label{tab1}
\begin{tabular*}{\linewidth}{@{\extracolsep{\fill}}lcccccccc}\hline\hline
               & ${{I_0}^M}$ & $I_0$    & $ \Delta(\%) $ & ${{V_0}^M}$ & $V_0$   & $ \Delta(\%)$ \\ \hline
Zero point     & 16.7262     & 16.6740   & 0.3131    & 17.3275     & 17.2011 & 0.7349    \\
$\delta$       & -0.3815     & -0.4127  & 7.5636   & -0.6711     & -0.4153 & 61.5775   \\
Slope          & -3.0540     & -2.9767 & 2.5969    & -2.9055     & -2.7623 & 5.1852    \\
$\delta$ Slope & 0.4283      & 0.4237   & 1.0751    & 0.7153      & 0.4393  & 62.8261   \\ \hline         
\end{tabular*}
\end{table*}

To analyze the impact in the PL relation when the median was used instead of the mean magnitude, comparisons between these two PL relations using $VI-$bands  were performed. First of all $\mathbf{\boldsymbol{I_0^M}}$ and  $\mathbf{\boldsymbol{I_0}}$ PL relations were compared employing the following statistical test whose details are discussed in Appendix A. 
The model used by the test was $\mathbf{\boldsymbol{I}}$=$\beta_1+\beta_2\ \boldsymbol{\delta}+\beta_3\ \boldsymbol{\delta_1}+\beta_4\ \boldsymbol{\log P}+\beta_5\ \boldsymbol{\delta \log P}+\beta_6\ \boldsymbol{\delta_1}\ \boldsymbol{\log P}+ \boldsymbol{\varepsilon}$, 
where $\boldsymbol{I}$ can be $\boldsymbol{{I_0}^M}$ or $\boldsymbol{I_0}$,  the test was conducted and the individual $p$-values for $\beta_3$ and $\beta_6$ were 0.0012 and 0.0120 respectively. The analysis shows that both variables are statistically significant in the model at a $5\%$ significance level,
meaning there are two different regressions with different zero points and slopes. 

The same test was used to compare the $\boldsymbol{{V_0}^M}$ and $\boldsymbol{V_0}$ PL relations, as a result, the model used by the test was: $\mathbf{\boldsymbol{V}}$=$\beta_1+\beta_2\ \boldsymbol{\delta}+\beta_3\ \boldsymbol{\delta_1}+\beta_4\ \boldsymbol{\log P}+\beta_5\ \boldsymbol{\delta \log P}+\beta_6\ \boldsymbol{\delta_1}\ \boldsymbol{\log P}+ \boldsymbol{\varepsilon}$. 
The test is significant at $5\%$ significance level for both parameters $\beta_3$ and $\beta_6$ because the $p$-values obtained are less than $0.0001$ and equal to $0.0015$ respectively, indicating that there are two different regression lines where $\boldsymbol{{V_0}^M}$ is used as dependent variable instead of $\boldsymbol{V_0}$. 
 It is worth saying that no problems were found in the two models used to perform these tests. 

Before measuring the impact when using the
median ($\mathbf{\boldsymbol{I_0^M}}$) magnitude instead of the intensity-mean 
($\mathbf{\boldsymbol{I_0}}$) magnitude commonly used,  it is %was 
important to consider the key features of these location measures.  The median is a robust statistical location measure with a high BDP of $50\%$, but it does not take into account the period while the mean magnitude does. 

After considering these location measures key features, the impact when using them to fit the PL relations, was calculated as the absolute percentage of change $\Delta (\%)$, representing the bias introduced by the median. Table \ref{tab1} reported these values.
The impact was huge in $V-$band due to the bias introduced for $\delta$ and $\delta$ Slope estimators that reached about $60\%$. 
Besides, analyzing the \textit{Ramsey-Reset} test for ${V_0}^M$, it was found that using the median instead of mean magnitude, changed the test statistic by $\sim$ $334\%$. 
The test statistic went from $-0.5215$ to $-2.2628$ when the PL relation was fitted with $V_0$ and ${V_0}^M$, respectively. 
This caused the test to remain significant even after improving the ${V_0}^M$ PL relation model, contrary to this model's residual plots which did not show any specification problems.

This analysis showed that the $I-$band was less affected than the $V-$band by the bias. 
When the median magnitude was used, the estimators for the adjustments introduced in the PL relation to solving the specification problems in $V-$band were more affected, as it is shown in Table \ref{tab1}. 
The bias could also affect the \textit{Ramsey-Reset} test since its test statistic had a huge increase as previously mentioned, which made this test still significant even after the new parameters were introduced to the model, in contrast to the model residual plots that did not show any problems.\\

\section{Summary and conclusions}

We analyzed the PL relation in $VI-$bands using data from OGLE-IV fundamental mode Cepheids in the LMC after correcting by extinction and discarding Cepheids with a higger extinction coefficient error.
The OLS regression commonly used is not appropriate to fit this relation because of influential points, that constitute an OLS assumptions violation.
Hence, the presence of influential Cepheids makes the robust $MM$-regression needed and we used it with a high BDP of $50\%$ to fit this relation instead of the commonly used OLS. 
Moreover, to ensure that heteroscedasticity was not a problem when fitting this robust regression, robust standard errors were used to guarantee a reliable method to estimate the PL relations using the median and mean magnitudes.  This analysis also allowed us to measure the impact of using the median.

The analysis of the PL relations in the VI-bands revealed specification problems in the model using statistical tests and the models' residual plots.
To solve the specification problems detected, we introduced two parameters in the model based on dummy variables to adjust the zero-point and slope when the Cepheids have periods larger than 10 days achieving it successfully. 
However, when analyzing the PL relation using the median $\boldsymbol{{V_0}^M}$ the specification problems persisted accordingly to the specification test but were not confirmed by the model residual plots. 
Besides, the bias in the adjustments to zero-point and slope could also affect the \textit{Ramsey-Reset test}. Based on the residual plots, the specification problems were solved and allowed us to reconfirm that there is a break in the PL relation when the fundamental mode Cepheids have a period larger than 10 days. 
This break was observed using the median and the mean magnitudes in both analyzed bands changing the zero point and the slope of the PL relation.

The comparisons between the PL relations showed enough evidence that regressions were not the same when the median was used instead of the mean magnitude, having different zero points and slopes. 
As a result, taking into account that the mean magnitude is robust and also considers the Cepheid phase, the differences in the model estimators were calculated as a bias introduced by using the median magnitude since this robust statistical location measure does not contemplate the phase. 

We recommend the use of the slope and zero point obtained by the $MM$-regressions with $50\%$ of BDP with robust standard errors and robust coefficient of determination to fit PL relation using the mean magnitudes in $VI-$bands, Equations (\ref{eq11}) and (\ref{eq13}). These values were obtained using OGLE-IV Cepheid data and deredenned using the Skowron extinction maps. It makes our PL relation appropriate to measure distances accurately, considering the break at 10 days. We also recommend the use of mean magnitudes instead of median magnitudes, because the first are robust
to outliers, making this location more precise.  The median magnitudes, on the other hand, despite being also robust do not take into account the phase of each Cepheid, introducing an important bias, especially in the $V-$band, affecting the PL relation model parameter estimators. Finally, it is important to perform a more detailed analysis to determine as precisely as possible the value of the period in which the PL relation breaks. We plan to do this in a future article, studying if this break can be eliminated using other possible measures of central tendency, different from those examined in the present work. \\

\section{acknowledgments}
The authors would like to thank the Vice Presidency of Research \& Creation’s Publication Fund at Universidad de los Andes for its financial support and also the Fondo de Investigaciones de la Facultad de Ciencias de la Universidad de los Andes, Colombia, through its Programa de Investigación c\'{o}digo INV-2023-162-2853.
Jaime Mu\~noz acknowledges financial support from Facultad de Ciencias and Facultad de Ingenier{\'i}a de la Universidad de los Andes, Colombia.  
\mbox{Santiago} Henao acknowledges the financial support given by the Departamento de Física de la Universidad de los Andes, in the form of a tutoring graduate assistance.
\vspace{1cm}

\bibliographystyle{aasjournal}

\bibliography{references}

\appendix

\section{Tests and other techniques used in the linear regression analysis}

In this section, the hypothesis testing and other techniques used in this PL relation analysis are explained. 
It is important to say that Statistically speaking when talking about hypothesis testing the null hypothesis is designated as $H_0$ and the alternative as $H_1$ or $H_a$. 

\textbf{Robust slope test}:  This test is used in linear regression to check the model's global significance. 
 This test contrasts two robust linear models: the full one with all the variables and the zero-point included (all parameters of the model) and the restricted one that only has the zero-point and no variables at all. 
 The hypotheses can be written as $H_0:$ all variable coefficients are equal to 0 vs $H_a:$ at least one variable coefficient is not equal to 0. 
 It is important to emphasize that the zero-point coefficient does not correspond to a variable, it is a model parameter. 
 A linear model with $k$
variables has $k+1$ parameters. 
If this test is not significant, there is enough evidence that the model is not overall adequate since there is no evidence to reject $H_0$. 

\textbf{Ramsey-Reset test's Godfrey and Orme variant}: This is a specification test to validate if the model is specified correctly with the needed variables and with the correct functional form. 
This test aims to contrast two hypotheses: $H_0$ which suggests there are no specification issues in the model, vs $H_a$ which states that there are specification problems in the model. 
These hypotheses contrast used in the context of this analysis the auxiliary model $\mathbf{\boldsymbol{y}}=\beta_1+\beta_2\ \protect\boldsymbol{\log P}+\alpha_1\ \mathbf{\boldsymbol{\widehat{y}^2}} +\boldsymbol{\varepsilon}$ where the model squared fitted values $\mathbf{\boldsymbol{\widehat{y}^2}}$ is added to build the auxiliary model and $\boldsymbol{y}$ could be in the context of this analysis:  $\mathbf{\boldsymbol{I_0^M}}$, $\mathbf{\boldsymbol{I_0}}$, $\mathbf{\boldsymbol{V_0^M}}$, or $\mathbf{\boldsymbol{V_0}}$. 
If this new variable is significant the test shows that there is enough evidence of specification problems. 
 It is worth emphasizing that only the significance analyzed is the one for the added variable so the null and alternative hypotheses of this test can be written as $H_0: \alpha_1=0$ and $H_a: \alpha_1\neq{0}$, respectively. 

\textbf{Fiting two linear relations in a single Equation}:  To fit two PL relations in a single Equation a dummy variable ($\delta$) is used. 
 This variable was included in the model to incorporate an adjustment in the original models to solve the specification problem caused by Cepheids with periods of 10 days or longer. 
Therefore, $\delta=0$ for Cepheids with periods of less than 10 days otherwise $\delta=1$. 
 Hence, the model fitted was $\mathbf{\boldsymbol{y}}=\beta_1+\beta_2\ \boldsymbol{\delta}+\beta_3\ \boldsymbol{{\log P}+\beta_4\ \boldsymbol{\delta}\ \boldsymbol{\log P}}+\boldsymbol{\varepsilon}$ where  $\mathbf{\boldsymbol{y}}$, in the context of this analysis, can be  $\mathbf{\boldsymbol{I_0^M}}$, $\mathbf{\boldsymbol{I_0}}$, $\mathbf{\boldsymbol{V_0^M}}$, or $\mathbf{\boldsymbol{V_0}}$. 
 For Cepheids with a period of less than 10 days, the fitted model becomes: 
$\mathbf{\boldsymbol{y}}=\beta_1+\beta_3\ \boldsymbol{\log P}+\boldsymbol{\varepsilon}$ because $\delta=0$, and 
$\mathbf{\boldsymbol{y}}=(\beta_1+\beta_2)\ + (\beta_3\ +\beta_4)\ \boldsymbol{\log P}+\boldsymbol{\varepsilon}$ when $\delta=1$ where $(\beta_1+\beta_2)$ are the model parameters related to zero point and $(\beta_3\ +\beta_4)$ to the slope.

\textbf{Test to compare the Median and Mean PL relations on each band}: To look for significant differences when the PL relation was fitted using the median magnitude and the mean magnitude as dependent variables, the following joint auxiliary model full model (with all variables) was fitted: $\mathbf{\boldsymbol{y}}$=$\beta_1+\beta_2\ \boldsymbol{\delta}+\beta_3\ \boldsymbol{\delta_1}+\beta_4\ \boldsymbol{\log P}+\beta_5\ \boldsymbol{\delta \log P}+\beta_6\ \boldsymbol{\delta_1}\ \boldsymbol{\log P}+ \boldsymbol{\varepsilon}$. 
 In the last Equation $\mathbf{\boldsymbol{y}}$ could be the median or the mean magnitude depending on the value taken by the $\delta_1-$dummy variable. 
When $\delta_1=1$, the median of the magnitude distribution was used as the dependent variable, and when $\delta_1=0$, the mean was used as the dependent variable. 
It is important to emphasize that the $\delta$ came from the fitted models for the PL relation and was used to introduce adjustments in the zero-point and the slope of the model when $\log P \geq{1}$ as explained before. 
 As a result, this test contrasted the following two hypotheses $H_0: \beta_3 = \beta_6 = 0$ vs $H_1$: At least one of them is not equal to 0 (notice that these were the parameters accompanying $\delta_1$). 
 This test used two models fitted in a single Equation, a technique already explained above, one for the median magnitudes and the other for the mean magnitudes. 
The purpose of this test was to identify any changes in the regression when the response variable $\mathbf{\boldsymbol{y}}$ changed between the median and mean magnitudes. 
 Therefore, the dataset used for this test had two observations for each Cepheid one with $\mathbf{\boldsymbol{y}}$ being median magnitude and $\delta_1 = 1$ while the other being the mean and $\delta_1=0$. The remaining variables remained identical for both observations.
 
\section{Additional Plots}
This section has plots of the linear regression analysis of the PL relation.

\begin{figure}[!htb]
    \begin{subfigure}{0.49\textwidth}
        \centering
        \includegraphics[width=\linewidth]{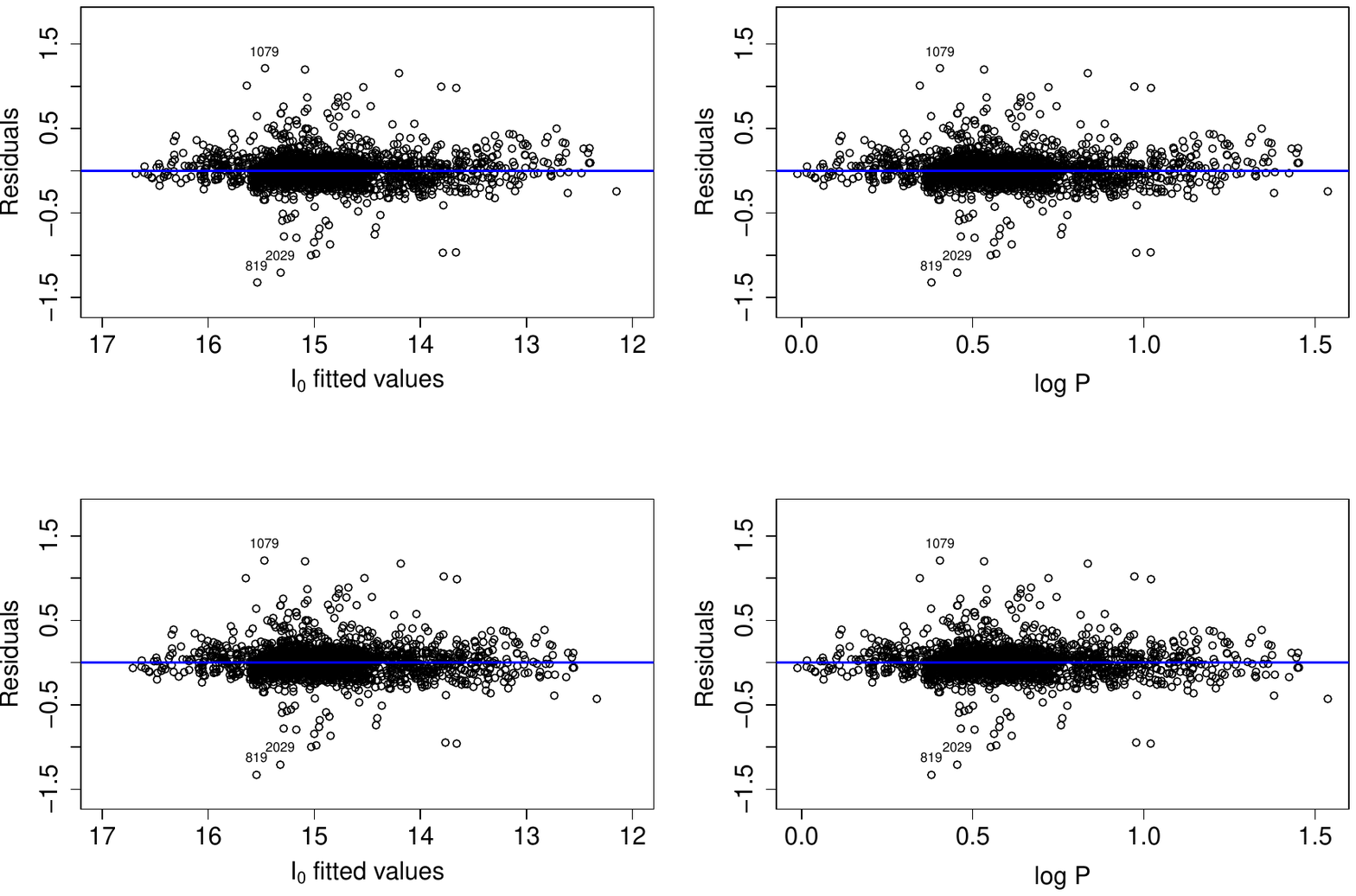}
        \caption{Residual plots}
        \label{fig:3a}
    \end{subfigure}%
    \begin{subfigure}{0.49\textwidth}
        \centering
        \includegraphics[width=\linewidth]{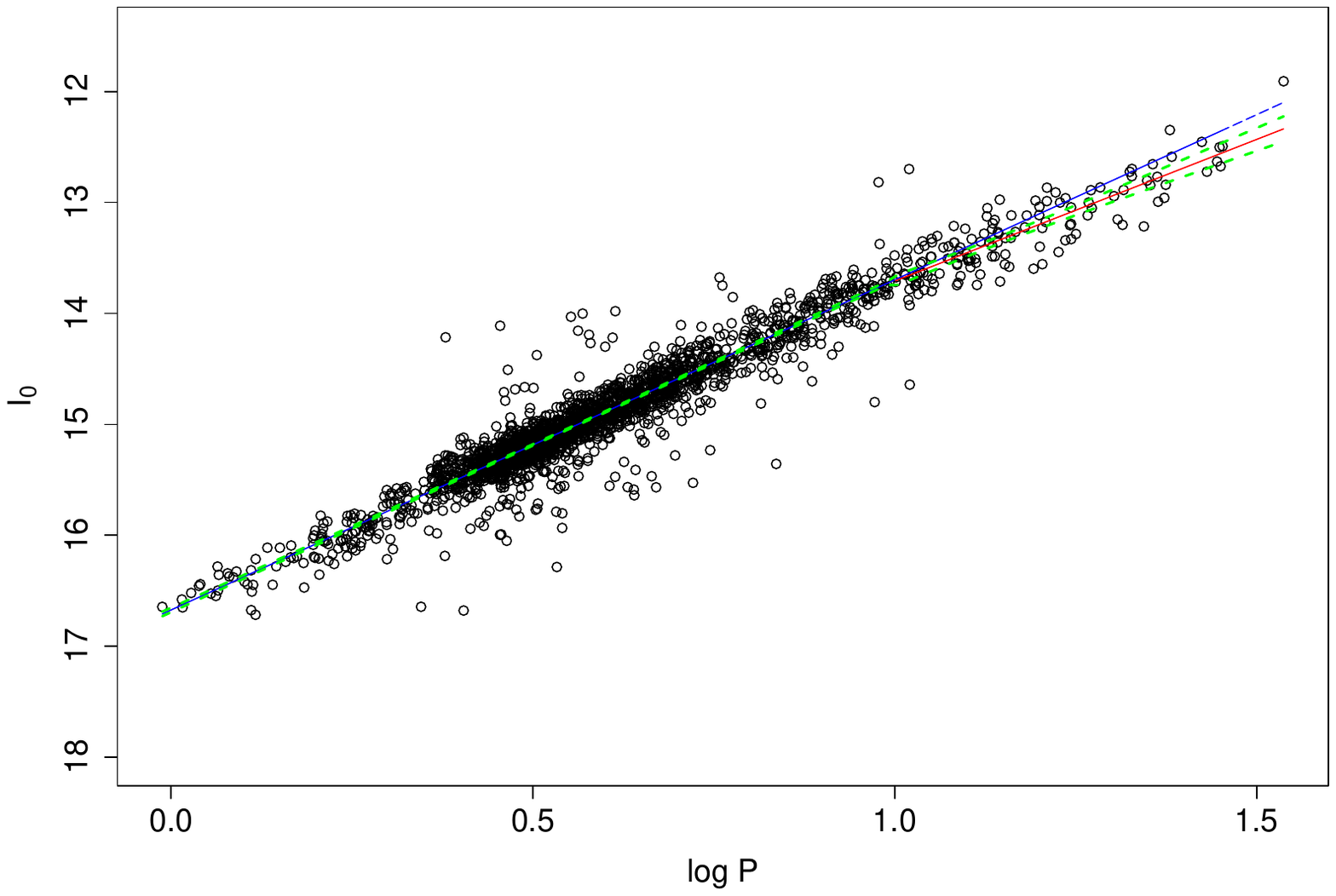}
        \caption{PL relation fitted}
        \label{fig:3b}
    \end{subfigure}
    \caption{PL relation fitted in $\mathbf{\boldsymbol{I_0}}$ by $MM$-regression. Plot (a): Checking the model residuals implies statistically plotting the residuals vs the fitted values (upper left panel) and the residuals vs the logarithm of the period (upper right panel). The first graph depicts a change in the residual model trend for magnitudes ranging from 13.3 to about 12.  In addition, the second plot shows a break in the residuals when $\log \ P \geq{1}$, 
which is evident for residuals above the blue trend line. 
This behavior indicates a shift in tendency in tendency when Cepheids have periods longer or equal to 10 days.
On the contrary,  when comparing the upper panel plots and bottom panels, the lasts show the residuals after the new parameters are introduced to the model to adjust the zero point and slope for Cepheids with periods longer or equal to 10 days, solving the specification problem since no break or change in tendency is shown by these plots. 
 Observations 819, 1079, and 2029 have the highest residuals and correspond to Cepheid stars OGLE-LMC-CEP-1583, OGLE-LMC-CEP-2162, and OGLE-LMC-CEP-4023 respectively. Plot (b): The regression given by Equation \eqref{eq11} is shown in a blue line for Cepheids with a $\log \ P<1$ and in a red line for Cepheids with a $\log \ P \geq{1}$. 
The blue line has been projected further with $\log \ P \geq{1}$ to enhance the visibility of the change when compared to the red line. 
 The area between the green dashed lines shows the $95\%$ confidence interval for $\mathbf{\boldsymbol{I_0}}$ expected value. 
This area increases when $\log \ P \geq{1}$ because Cepheids having these periods are more scattered around the regression line. }
    \label{fig:3}
\end{figure}

\begin{figure}[!htb]
    \begin{subfigure}{0.49\textwidth}
        \centering
        \includegraphics[width=\linewidth]{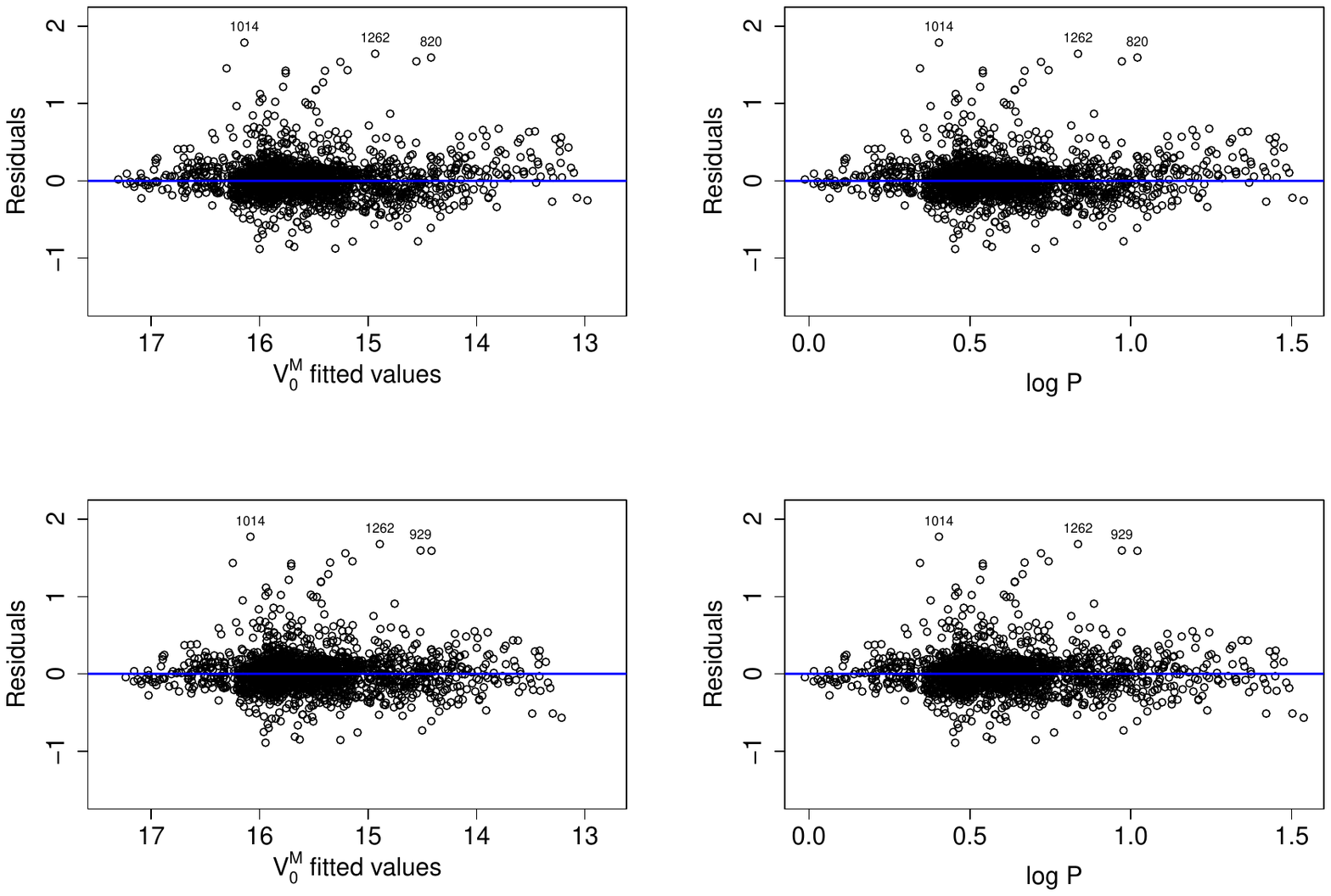}
        \caption{Residual plots}
        \label{fig:4a}
    \end{subfigure}%
    \begin{subfigure}{0.49\textwidth}
        \centering
        \includegraphics[width=\linewidth]{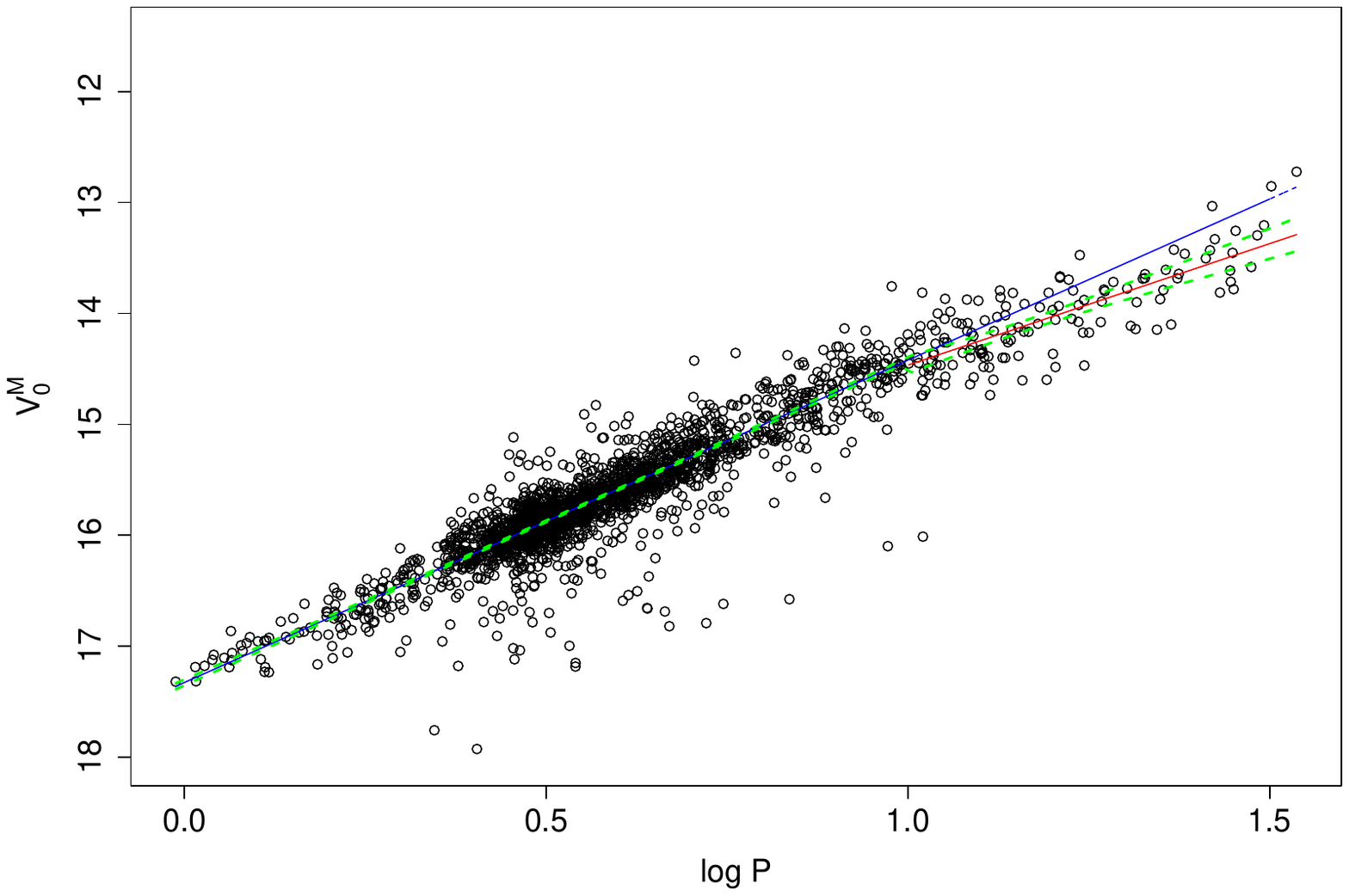}
        \caption{PL relation fitted}
        \label{fig:4b}
    \end{subfigure}
    \caption{PL relation fitted in $\mathbf{\boldsymbol{V_0}^M}$ by $MM$-regression. Plot (a): Checking the model residuals implies statistically plotting the residuals vs the fitted values (upper left panel) and the residuals vs the logarithm of the period (upper right panel). The first graph depicts a change in the residual model trend for magnitudes ranging from 13.3 to about 12.  In addition, the second plot shows a break in the residuals when $\log \ P \geq{1}$, 
which is evident for residuals above the blue trend line. 
This behavior indicates a shift in tendency when Cepheids have periods longer or equal to 10 days.
On the contrary,  when comparing the upper panel plots and bottom panels, the lasts show the residuals after the new parameters are introduced to the model to adjust the zero point and slope for Cepheids with periods longer or equal to 10 days, solving the specification problem since no break or change in tendency is shown by these plots. 
 Observations 820, 929 1014, and 1262 have the highest residuals and correspond to Cepheid OGLE-LMC-CEP-1719, LMC-CEP-1940,
OGLE-LMC-CEP-2162, and OGLE-LMC-CEP-2675 respectively. Plot (b): The regression given by Equation \eqref{eq12} is shown in a blue line for Cepheids with a $\log \ P<1$ and in a red line for Cepheids with a $\log \ P \geq{1}$. 
The blue line has been projected further with $\log \ P \geq{1}$ to enhance the visibility of the change when compared to the red line. 
 The area between the green dashed lines shows the $95\%$ confidence interval for $\mathbf{\boldsymbol{V_0}^M}$ expected value. 
This area increases when $\log \ P \geq{1}$ because Cepheids having these periods are more scattered around the regression line. }
    \label{fig:4}
\end{figure}

\begin{figure}[!htb]
    \begin{subfigure}{0.49\textwidth}
        \centering
        \includegraphics[width=\linewidth]{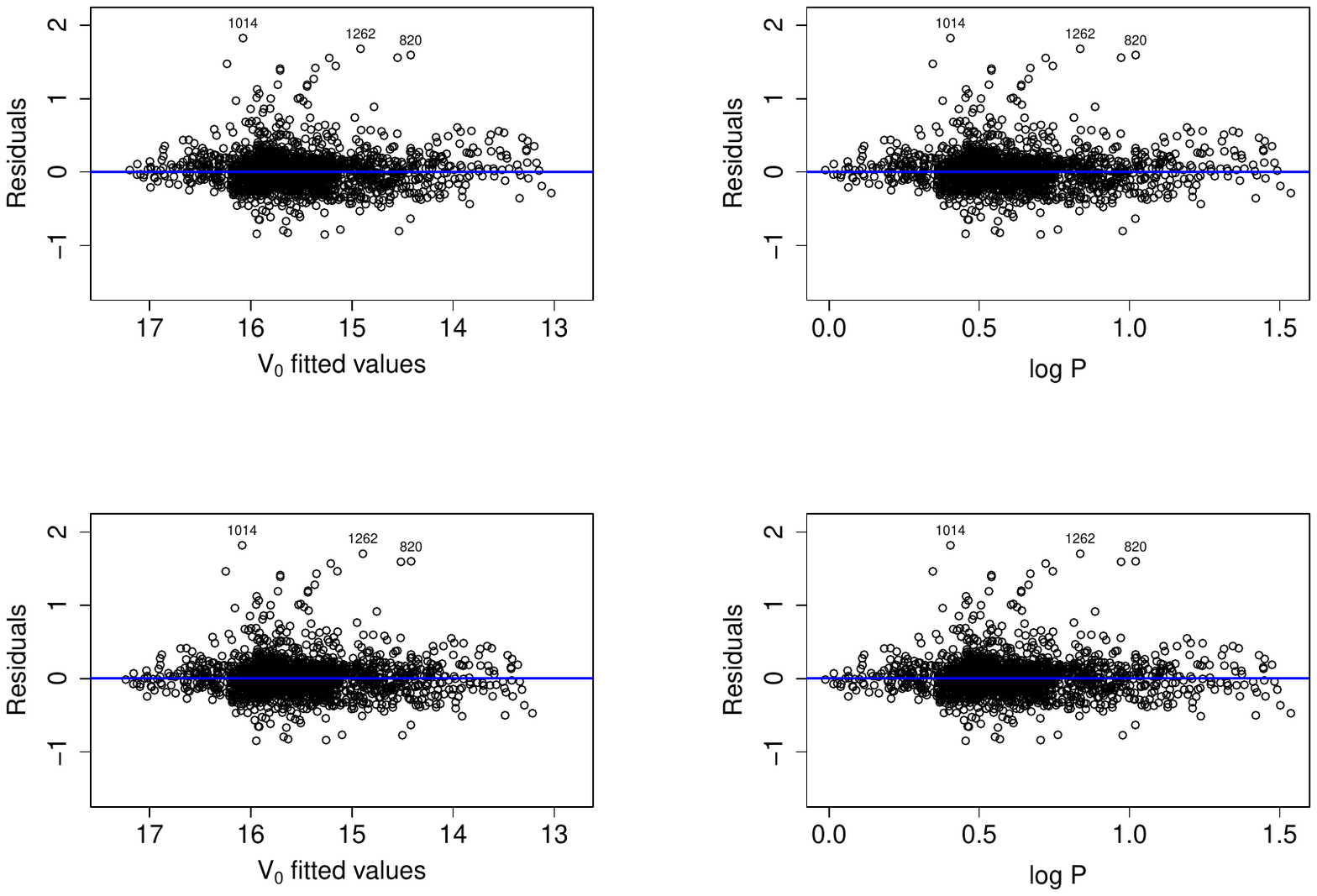}
        \caption{Residual plots}
        \label{fig:5a}
    \end{subfigure}%
    \begin{subfigure}{0.49\textwidth}
        \centering
        \includegraphics[width=\linewidth]{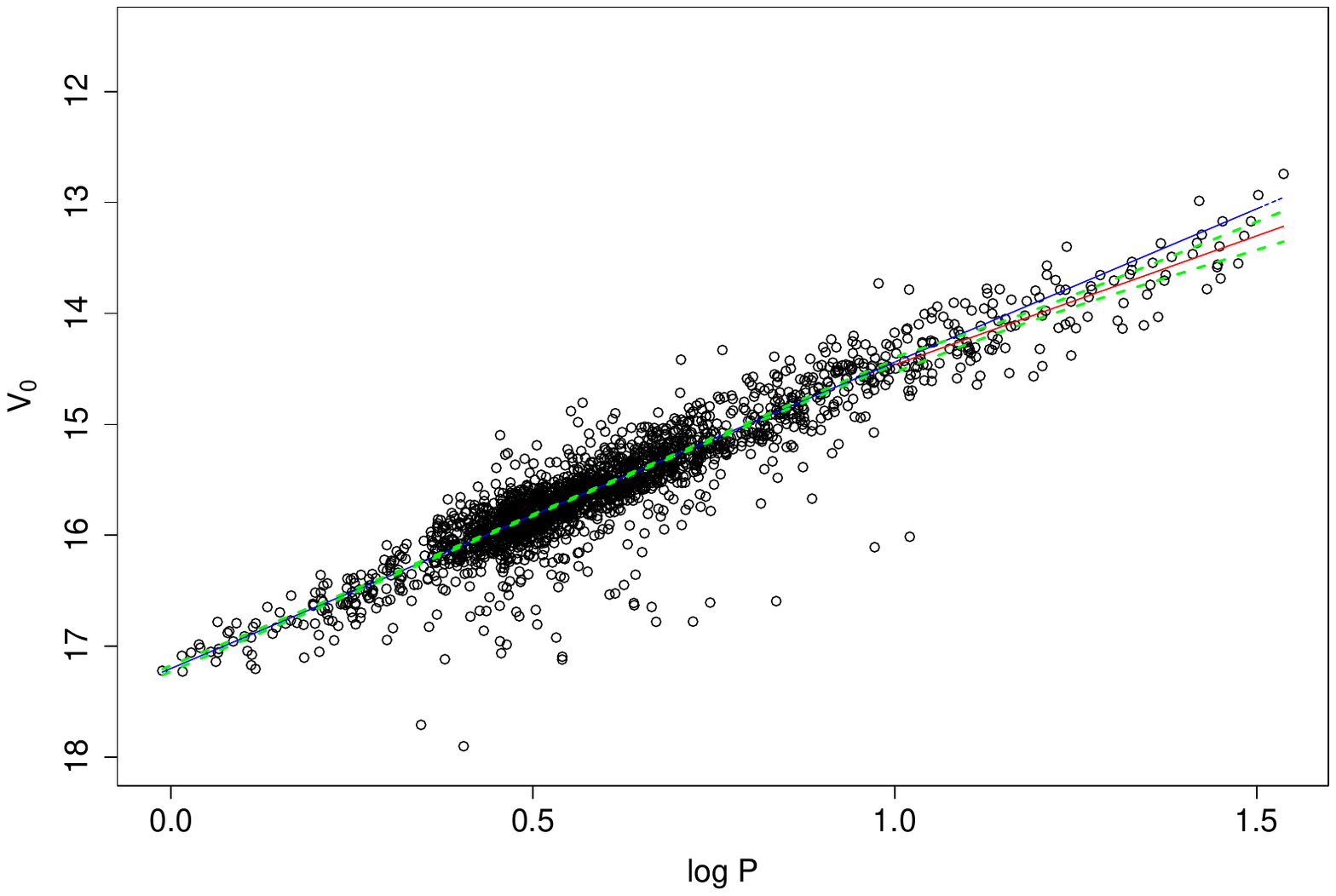}
        \caption{PL relation fitted}
        \label{fig:5b}
    \end{subfigure}
    \caption{PL relation fitted in $\mathbf{\boldsymbol{V_0}}$ by $MM$-regression. Plot (a): Checking the model residuals implies statistically plotting the residuals vs the fitted values (upper left panel) and the residuals vs the logarithm of the period (upper right panel). The first graph depicts a change in the residual model trend for magnitudes ranging from 13.3 to about 12.  In addition, the second plot shows a break in the residuals when $\log \ P \geq{1}$, 
which is evident for residuals above the blue trend line. 
This behavior indicates a shift in tendency when Cepheids have periods longer or equal to 10 days.
On the contrary,  when comparing the upper panel plots and bottom panels, the lasts show the residuals after the new parameters are introduced to the model to adjust the zero point and slope for Cepheids with periods longer or equal to 10 days, solving the specification problem since no break or change in tendency is shown by these plots. 
 Observations 820, 1014, and 1262 have the highest residuals and correspond to Cepheid OGLE-LMC-CEP-1719, 	
OGLE-LMC-CEP-2162, and OGLE-LMC-CEP-2675 respectively. Plot (b): The regression given by Equation \eqref{eq13} is shown in a blue line for Cepheids with a $\log \ P<1$ and in a red line for Cepheids with a $\log \ P \geq{1}$. 
The blue line has been projected further with $\log \ P \geq{1}$ to enhance the visibility of the change when compared to the red line. 
 The area between the green dashed lines shows the $95\%$ confidence interval for $\mathbf{\boldsymbol{V_0}}$ expected value. 
This area increases when $\log \ P \geq{1}$ because Cepheids having these periods are more scattered around the regression line. }
    \label{fig:5}
\end{figure}

\end{document}